\documentclass{llncs}
\usepackage[latin1]{inputenc}
\usepackage{amssymb}

\usepackage[all]{xy}

\usepackage{theorem}
\newtheorem{theo}{Theorem}[section]
\newtheorem{prop}[theo]{Proposition}

\theorembodyfont{\normalfont}

\newtheorem{defi}[theo]{Definition}
\newtheorem{exam}[theo]{Example}

\newcommand{\signature}{\Omega}  
\newcommand{\features}{\mathcal{F}}  
\newcommand{\nodes}{\mathcal{N}} 
\newcommand{\edges}{\mathcal{E}} 
\newcommand{\nlabel}{\mathcal{L}^n} 
\newcommand{\elabel}{\mathcal{L}^e} 
\newcommand{\source}{\mathcal{S}}  
\newcommand{\target}{\mathcal{T}}  
\newcommand{\var}{\bullet}
\newcommand{\field}{\Rightarrow}
\newcommand{\lb}{\!:\!}
\newcommand{\lr}{\gg} 
\newcommand{\gr}{\gg} 

\title{A modal Logic for Termgraph Rewriting \thanks{
This work has been partly funded by the project ARROWS of
the French \it{Agence Nationale de la Recherche}.}}

\author{Ph. Balbiani\inst{1} \and R. Echahed\inst{2} \and A. Herzig\inst{1}}

\institute{ Université de Toulouse, CNRS \\
            Institut de Recherche en Informatique de Toulouse (IRIT)\\
            118 route de Narbonne,
            31062 Toulouse Cedex 9, France \\
           \email{$\{$balbiani,herzig$\}$@irit.fr}
            \and
            Laboratoire LIG\\
            Bât IMAG C \\
            BP 53 \\
            38041 Grenoble Cedex, France \\
            \email{Rachid.Echahed@imag.fr}}

\date{November 3, 2008}

\begin{document}

\maketitle

\begin{abstract}
We propose a modal logic tailored to describe graph transformations
and discuss some of its properties.  We focus on a particular class of
graphs called termgraphs. They are first-order terms augmented with
sharing and cycles. Termgraphs allow one to describe classical
data-structures (possibly with pointers) such as doubly-linked lists,
circular lists etc. We show how the proposed logic can faithfully
describe (i) termgraphs as well as (ii) the application of a termgraph
rewrite rule (i.e. matching and replacement) and (iii) the computation
of normal forms with respect to a given rewrite system. We also show
how the proposed logic, which is more expressive than propositional
dynamic logic, can be used to specify shapes of classical
data-structures (e.g. binary trees, circular lists etc.).
\end{abstract}
\section{Introduction}

Graphs are common structures widely used in several areas in
computer science and discrete mathematics.  Their transformation
constitute a domain of research per se with a large number of
potential applications \cite{handbook1,handbook2,handbook3}. There are many different ways
to define graphs and graph transformation. We consider in this paper
structures known as \emph{termgraphs} and their transformation via
rewrite rules \cite{BVG87,Plu98a}.  Roughly speaking, a termgraph is a
first-order term with possible sharing (of sub-terms) and cycles. Below we depict three examples of termgraphs : $G_0$ is a classical first-order term. $G_1$ represents the same expression as $G_0$ but argument $x$ is shared. $G_1$ is often used to define the function double $double(x) = G_1$. The second termgraph $G_2$ represents a circular list of two ``records'' (represented here by operator $cons$) sharing the same content $G_1$.

$$ 
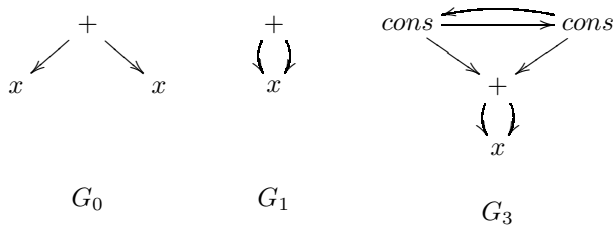
\xymatrix@R=1pc@C=1pc
     {& +\ar[dl]\ar[dr]   \\
       x & & x \\
       & \\
& G_0
} \hspace{1cm}
 \xymatrix@R=1pc@C=1pc
     {+\ar@/_/[d]\ar@/^/[d]   \\
       x  \\
         \\
      G_1
}\hspace{1cm}
\xymatrix@R=1pc@C=1pc
     {    cons \ar[dr]\ar[rr] & & cons \ar[dl] \ar@/_/[ll] \\
    & + \ar@/_/[d]\ar@/^/[d]  &  \\
      & x   & \\
    & G_3
}$$

Termgraphs allow to represent real-world data structures (with pointers) such as
circular lists, doubly-linked lists etc \cite{Echahed08},
and rewriting allows to efficiently process such graphs.
They are thus a suitable framework for declarative languages dealing
with such complex data structures.
However, while there exist rewriting-based proof methods
for first-order terms,
there is a lack of appropriate termgraph rewriting proof methods,
diminishing thus their operational benefits.
Indeed, equational logic provides a logical setting
for first-order term rewriting \cite{baader1998}, and many theorem provers use
rewrite techniques in order to efficiently achieve equational reasoning.
In \cite{CEP08} an extension of first-order (clausal)
logic dealing with termgraphs has been proposed to give a logic
counterpart of termgraph rewriting.
In such a logic operations are interpreted as continuous functions \cite{Tiu78,Tiu79} and
bisimilar graphs cannot be distinguished (two termgraphs are bisimilar
if and only if they represent the same rational term).
Due to that, reasoning on termgraphs is unfortunately much trickier
than in first-order classical logic.
For example, equational theories on termgraphs are not recursively enumerable
whereas equational theories on terms are r.e.).

In this paper, we investigate a modal logic with possible worlds semantics
which better fits the operational features of termgraph rewriting systems.
Termgraphs can easily be interpreted within the framework of possible
worlds semantics, where nodes are considered as worlds and edges as modalities.
Based on this observation, we investigate a new modal logic
which has been tailored to fit termgraph rewriting.  We show how
termgraphs as well as rewrite rules can be specified by means of modal
formulae.
In particular we show how a rewrite step can be defined by means
of a modal formula which encodes termgraph matching (graph homomorphism) and
termgraph replacement (graph construction and modification). We show also how to
define properties on such structures,
such as being a list, a circular list, a tree, a binary
tree.
The computation of termgraph normal form is formulated in this new logic.
In addition, we formulate invariant preservation by rewriting
rules and discuss subclasses for which validity is decidable.

The next two sections introduce respectively the considered class of
termgraph rewrite systems and the proposed modal logic.
In section~\ref{sec-definability} we discuss briefly the expressive power
of the modal logic and show particularly how graph homomorphisms can
be encoded.  In section~\ref{sec-logic-transformation} we show how
elementary graph transformations can be expressed as modal logic
formulae whareas section~\ref{sec-traduction} shows how termgraph
rewriting can be specified as modal formulae.
Section~\ref{conclusion} gives some concluding remarks.

\section{Termgraph Rewriting}
\label{sec:graph-rew}

This section defines the framework of graph rewrite systems that we
consider in the paper. There are different approaches in the
literature to define graph transformations. We follow here an algorithmic
approach to termgraph rewriting \cite{BVG87}. Our definitions are consistent
with \cite{Echahed08}.

\begin{defi}[Graph]
\label{defi:graph-graph}
\ \\
A \emph{termgraph}, or simply a \emph{graph} is a tuple $G = (\nodes,\edges,\nlabel,\elabel,\source, \target)$ which consists of
a finite set of nodes $\nodes$,
a finite set of edges $\edges$,
a (partial) node labelling function  $\nlabel : \nodes \to \signature$  which associates labels
in $\signature$ to  nodes in $\nodes$,
a (total) edge labelling function  $\elabel : \edges \to \features$
which associates, to every edge in
$\edges$, a label (or feature) in $\features$,
a source function $\source : \edges \to \nodes$ and
a target function $\target : \edges \to \nodes$
which specify respectively, for every edge $e$, its source
$\source(e)$ and its target $\target(e)$.

Note that $G$ is a first-order term if and only if $G$ is a tree.

We assume that the labelling of edges $\elabel$ fulfills the
following additional \emph{determinism} condition: $\forall e_1, e_2 \in
\edges,(\source(e_1) = \source(e_2)
\mbox{ and } \elabel(e_1) = \elabel(e_2)) \mbox{
implies } e_1 = e_2$.
This last condition expresses the fact that for every node $n$
there exists at most one edge $e$ of label $a$
such that the source of $e$ is $n$.
We denote such an edge by the tuple $(n,a,m)$ where $m$ is the target of edge $e$.
\end{defi}

\noindent
\textbf{Notation}: For each labelled node $n$ the fact that $\omega=\nlabel(n)$
is written $n \lb \omega$, and each unlabelled node $n$ is written as
$n \lb \var$.
This `unlabelled' symbol $\var$ is used in termgraphs to represent
anonymous variables.
$n \lb \omega (a_1 \field n_1, \ldots,a_k \field n_k)$
describes a node $n$ labelled by symbol $\omega$ with $k$ outgoing edges,
$e_1, \ldots, e_k$, such that for every edge $e_i$,
$\elabel(e_i) = a_i$, $\source(e_i) = n$ and $\target(e_i) = n_i$.
In the sequel we will use the linear notation of termgraphs \cite{BVG87}
defined by the following grammar.
The variable $A$ (resp.~ $F$ and $n$)
ranges over the set $\signature$ (resp.~$\features$ and $\nodes$)~: \\
\begin{tabular}{lll}
{\sc TermGraph} & ::= & {\sc Node} $\mid$ {\sc Node} {\bf +} {\sc TermGraph} \\
{\sc Node}  & ::= & $n$:$A$({F $\field$ \sc Node},\ldots,{F $\field$ \sc Node})
                   $\mid$ $n$:$\bullet$ $\mid$ $n$
\end{tabular} \\
the operator ${\bf +}$ stands for the disjoint
union of termgraph
definitions. We assume that every node is labelled at most once.  The
expression $n \lb \omega (n_1, \dots, n_k)$ stands for $n \lb \omega
(1 \field n_1, \dots, k \field n_k)$.

A \emph{graph homomorphism},  $h:G \to G_1$, where
$G=(\nodes,\edges,\nlabel,\elabel,\source, \target)$
and $G_1=(\nodes_1,\edges_1,\nlabel_1,\elabel_1,\source_1,
\target_1)$ is a pair of functions $h =(h^n,h^e)$ with
 $h^n:\nodes \to \nodes_1$ and
$h^e:\edges \to \edges_1$
which preserves the
labelling of nodes and edges as well as the source and target
functions.
This means that
for each labelled node $m$ in $G$, $\nlabel_1(h^n(m))=\nlabel(m)$
and for each edge $f$ in $G$,
$\elabel_1(h^e(f))=\elabel(f)$, $\source_1(h^e(f)) =
h^n(\source(f))$ and $\target_1(h^e(f)) = h^n(\target(f))$.  Notice that the
image by $h^n$ of an unlabelled node may be any node.

\textbf{Remark}: Because of the determinism condition, a homomorphism $h:G \to G_1$ is completely defined by the function  $h^n:\nodes \to \nodes_1 $ which should satisfy the following conditions : for each
labelled node $m$ in $G$, $\nlabel_1(h^n(m))=\nlabel(m)$ and for every outgoing edge from $m$, say $(m,a,w)$, for some feature $a$ and node $w$, the edge $(h^n(m), a, h^n(w))$ belongs to $\edges_1$.

\begin{exam}

Let $B_1$, $B_2$ and  $B_3$  be the following termgraphs.

$$ B_1 : 
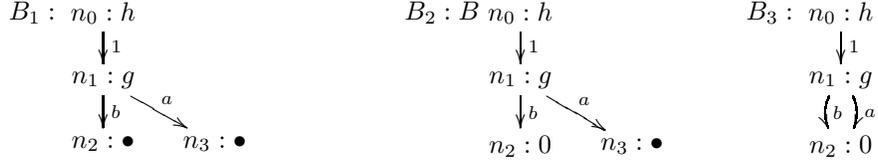
\xymatrix@R=1pc@C=1pc
     {  n_0 : h \ar^{1}[d] \\
        n_1:g \ar^{b}[d]\ar^{a}[dr] \\
       n_2:\bullet & n_3:\bullet
} \hspace{2cm}
B_2 :B \xymatrix@R=1pc@C=1pc
     {  n_0 : h \ar^{1}[d] \\
        n_1:g \ar^{b}[d]\ar^{a}[dr] \\
       n_2:0 & n_3:\bullet
} \hspace{1cm}
 B_3 : \xymatrix@R=1pc@C=1pc
     {  n_0 : h \ar^{1}[d] \\
        n_1:g \ar^{b}@/_/[d]\ar^{a}@/^/[d] \\
        n_2:0
}$$

and $h$ and $h'$ be two functions on nodes defined as follows:
$h(n_i) = n_i$ for $i$ in $\{0,1,2,3\}$ and
$h'(n_i) = n_i$ for $i$ in $\{0,1,2\}$ and $h'(n_3) = n_2$.
$h$ defines a homomorphism from $B_1$ to $B_2$.
$h'$ defines a homomorphism from  $B_1$ to $B_3$ and from $B_2$ to $B_3$.
There is no homomorphism from $B_3$ to $B_2$ or to $B_1$, nor from $B_2$ to $B_1$.

\end{exam}

The following definition introduces a notion of actions. Each action specifies an elementary transformation of graphs. These elementary  actions are used later on to define graph transformations by means of rewrite rules.

\begin{defi}[Actions]
An action has one of the following forms.

\begin{itemize}

\item a \textbf{node definition} or \textbf{node labelling}
$n \lb f(a_1 \field n_1, \ldots,a_k \field n_k)$
 where $n, n_1, \ldots, n_k$ are nodes and $f$ is a label of node
 $n$. For $i \in \{1, \ldots, k\}$, $a_i$ is the label of an edge,
 $e_i$, such that ($\elabel(e_i) = a_i$) and whose source is $n$
($\source(e_i) = n$) and target is node $n_i$ ($\target(e_i) = n_i$).
 This action, first creates a new node $n$ if $n$ does not already exist
in the context of application of the action. Then node $n$  is defined by its label and its outgoing edges.

\item an \textbf{edge redirection} or \textbf{local redirection} $n \lr_a m $

 where $n, m$ are nodes and $a$ is the feature of an edge $e$ outgoing
 node $n$ ($\source(e) = n$ and $\elabel(e) = a$).  This action is an
 edge redirection and means that the target of edge $e$ is redirected
 to point to the node $m$ (i.e., $\target(e) = m$ after performing the action
 $n \lr_a m $).

\item a \textbf{global redirection} $n \gr m$

 where $n$ and $m$ are nodes.  This means that all edges $e$ pointing
 to $n$ ($\target(e)= n$) are redirected to point to the node $m$
 ($\target(e)=m$).

\end{itemize}

The result of applying an action $\alpha$ to a termgraph
$G=(\nodes,\edges,\nlabel,\elabel,\source, \target)$ is denoted by
$\alpha[G]$ and is defined as the following termgraph
$G_1=(\nodes_1,\edges_1,\nlabel_1,\elabel_1,\source_1,\target_1)$ such
that~:

\begin{itemize}

\item If $\alpha = n \lb f(a_1 \field n_1, \ldots, a_k \field n_k)$
then

\begin{itemize}
\item
$\nodes_1 =  \nodes \cup \{ n, n_1, \ldots,n_k \}$,
\item
$\nlabel_1(n) = f$ and
$\nlabel_1(m) = \nlabel (m)$ if $m \not = n$,

\item
Let $E = \{e_i \mid 1 \leq i \leq k, e_i \mbox{ is an edge such that } \source(e_i)= n, \target(e_i)= n_i \mbox{ and } \elabel(e_i) = a_i \}$.
$\edges_1 = \edges \cup E $,
\item
$\elabel_1 (e) =\left \{
\begin{array}{l l}
a_i & \quad \mbox{ if }  e = e_i \in E \\
\elabel(e) & \quad \mbox{ if } e \not\in E \\
\end{array}\right. $

\item
$\source_1 (e) =\left \{
\begin{array}{l l}
n & \quad \mbox{ if }  e = e_i \in E \\
\source(e) & \quad \mbox{ if } e \not\in E \\
\end{array}\right. $

\item
$\target_1 (e) = \left \{
\begin{array}{l l}
n_i & \quad \mbox{ if }  e = e_i \in E \\
\target(e) & \quad \mbox{ if } e \not\in E \\
\end{array}\right. $
\end{itemize}
$\cup$ denotes classical union. This means that the nodes  in $\{ n, n_1, \ldots,n_k \}$ which already belong to $G$ are reused whereas the others are new.

\item If $\alpha = n  \lr_a m$  then
\begin{itemize}
\item
$\nodes_1 = \nodes$,
$\nlabel_1 = \nlabel$,
$\elabel_1 = \elabel$,
$\source_1 = \source$ and

\item
Let $e$ be the edge of label $a$  outgoing $n$. \\
$\target_1 (e) = m$ and $\target_1 (e') = \target(e')$ if $e' \not= e$.
\end{itemize}
 \item If $\alpha = n \gr m$ then
$\nodes_1 = \nodes$,
$\nlabel_1 = \nlabel$,
$\elabel_1 = \elabel$,
$\source_1 = \source$ and

\[\target_1 (e) = \left \{
\begin{array}{l l}
m & \quad \mbox{ if } \target(e)=n \\
\target(e) & \quad \mbox{ otherwise} \\
\end{array}\right. \]
\end{itemize}

A \emph{rooted} termgraph is a termgraph $G$ with a
distinguished node $n$ called its root. We write $G =(\nodes,\edges,\nlabel,\elabel,\source, \target, n)$.  The application of an action
$\alpha$ to a rooted termgraph $G = (\nodes,\edges,\nlabel,\elabel,\source, \target, n)$ is a rooted termgraph $G_1 =(\nodes_1,\edges_1,\nlabel_1,\elabel_1,\source_1, \target_1, n_1)$
such that $G_1 = \alpha[G]$ and root $n_1$ is defined as follows~:
\begin{itemize}
\item $n_1 = n$ if $\alpha$ is not of the form $n \gr p$.
\item $n_1 = p$ if $\alpha$ is  of the form $n \gr p$.
\end{itemize}
\end{defi}
The application of a sequence of actions $\Delta$ to a (rooted)
termgraph $G$ is defined inductively as follows~: $\Delta[G]= G$ if
$\Delta$ is the empty sequence and $\Delta[G] = \Delta'[\alpha[G]]$ if
$\Delta = \alpha;\Delta'$ where ``$;$'' is the concatenation (or sequential)
operation.  Let $h$ be a homomorphism. We denote by $h(\Delta)$ the
sequence of actions obtained from $\Delta$ by substituting every node
$m$ occurring in $\Delta$ by $h(m)$.

\begin{exam}
\label{example-actions}
This example  illustrates the application of actions.
Let $H_1$, $H_2$, $H_3$, $H_4$ and $H_5$ be the following termgraphs.

$$ H_1 : 
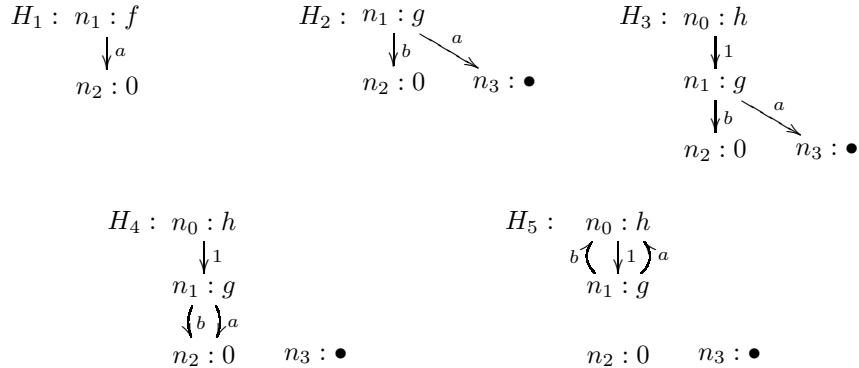
\xymatrix@R=1pc@C=1pc
     {     n_1:f\ar^{a}/_/:[d] \\
           n_2: 0
} \hspace{2cm}
H_2 : \xymatrix@R=1pc@C=1pc
     {  n_1:g \ar^{b}[d]\ar^{a}[dr] \\
        n_2:0 & n_3:\bullet
} \hspace{1cm}
 H_3 : \xymatrix@R=1pc@C=1pc
     {  n_0 : h \ar^{1}[d] \\
        n_1:g \ar^{b}[d]\ar^{a}[dr] \\
       n_2:0 & n_3:\bullet
}$$

$$ H_4 : \xymatrix@R=1pc@C=1pc
     {  n_0 : h \ar^{1}[d] \\
        n_1:g \ar^{b}@/_/[d]\ar^{a}@/^/[d] \\
        n_2:0 & n_3:\bullet
}\hspace{2cm}
H_5 : \xymatrix@R=1pc@C=1pc
     {  n_0 : h \ar^{1}[d] \\
        n_1:g \ar_{a}@/_1pc/[u]\ar^{b}@/^1pc/[u]\\
        n_2:0 & n_3:\bullet
}$$

$H_2$ is obtained from $H_1$ by applying the action $ n_1 \lb g(b
\field n_2, a \field n_3)$.  $n_1$ is relabelled whereas $n_3$ is a
new unlabelled node.  $H_3$ is obtained from $H_2$ by applying the
action $\alpha = n_0 \lb h(n_1)$. $n_0$ is a new node labelled by
$h$. $h$ has one argument $n_1$.  $H_4$ is obtained from $H_3$ by
applying the action $n_1 \gg_a n_2$. The effect of this action is to
change the target $n_3$ of the edge $(n_1, a, n_3)$ by $n_2$. $H_5$ is
obtained from $H_4$ by applying the action $n_2 \gg n_0$. This action
redirects the incoming edges of node $n_2$ to target node $n_0$.

\end{exam}

\begin{defi}[Rule, system, rewrite step]

A \emph{rewrite rule} is an expression of the
form $l  \to r$ where  $l$ is a termgraph and $r$ is a sequence of actions.
A rule is written $l \to (a_1 ,\ldots, a_n )$ or $l \to a_1 ;\ldots; a_n $
where the $a_i 's$ are elementary actions.
A \emph{termgraph rewrite system} is a set of rewrite rules.
We say that the term-graph $G$ rewrites to $G_1$
using the rule $l \to r$ iff there exists a homomorphism $h: l \to G$
 and $G_1=h(r)[G]$. We write
$G \to_{l \to r} G_1$,  or simply $G \to G_1$.
\end{defi}

\begin{exam}
We give here an example of a rewrite step.
Consider the following rewrite rule:
 $$n_1 : g(a \Rightarrow n_2 : \bullet, b \Rightarrow n_3: \bullet) \to n_0 : h(1 \Rightarrow n_1) ; n_1 \gg_a n_2 ; n_2 \gg n_0 $$

The reader may easily verify that the graph $H_2$ of
Example~\ref{example-actions} can be rewritten by the considered rule
into the graph $H_5$ of Example~\ref{example-actions}.

\end{exam}

\begin{exam}
\label{tGRSexamples}
We give here somme illustrating examples of the considered class of rewrite systems.
We first define an operation, $insert$, which inserts an element in a circular list.
\noindent
\begin{tabbing}
$r:insert(m:\bullet, p_1:cons(m_1 : \bullet, p_2)) {\bf +}  p_3:cons(m_2, p_1) \to$ \= $ p_4 : cons(m,p_1); p_3 \lr_2 p_4 ; r \gr p_4 $ \kill
$r:insert(m:\bullet, p_1:cons(m_1 : \bullet, p_1)) \to p_2 : cons(m, p_1); p_1 \lr_2 p_2 ; r \gr p_2 $ \\
$r:insert(m:\bullet, p_1:cons(m_1 : \bullet, p_2)) {\bf +}  p_3:cons(m_2, p_1) \to p_4 : cons(m,p_1);$ \\ \> $p_3 \lr_2 p_4 ; r \gr p_4 $ \\
\end{tabbing}
As a second example, we define below the operation $ length$  which
computes the number of elements of any, possibly circular, list.

\noindent
$r:length(p:\bullet) \rightarrow r':length'(p,p) ; r \gg r'$\\
$r:length'(p_1 : nil, p_2 : \bullet) \rightarrow r': 0; r \gg r'$\\
$r:length'(p_1 : cons(n:\bullet, p_2 : \bullet), p_2) \rightarrow r':succ(0) ; r \gg r'$\\
$r:length'(p_1 :cons(n:\bullet, p_2 : \bullet), p_3 : \bullet) \rightarrow  r':s(q: \bullet) ; q:length'(p_2 , p_3) ; r \gg r'$\\

Pointers help very often to enhance the efficiency of algorithms. In
the following, we define the operation $reverse$ which performs the
so-called ``in-situ list reversal''.
\noindent
\begin{tabbing}
$o:reverse'(p_1 :cons(n : \bullet, p_2:cons(m : \bullet,p_3 : \bullet), p_4 : \bullet) \rightarrow $ \= $ p_{1} \gg_2 p_4 ; o \gg_1 p_2 ; o \gg_2 p_1 $ \kill
$o:reverse(p:\bullet) \rightarrow o':reverse'(p, q:nil); o \gg o'$ \\
$o:reverse'(p_1 : cons(n : \bullet, q : nil), p_2 : \bullet) \rightarrow p_{1} \gg_2 p_2 ; o \gg p_1 $ \\
$o:reverse'(p_1 :cons(n : \bullet, p_2:cons(m : \bullet,p_3 : \bullet), p_4 : \bullet) \rightarrow p_{1} \gg_2 p_4 ; $ \\ \> $o \gg_1 p_2 ; o \gg_2 p_1 $\\
\end{tabbing}

\noindent

The last example illustrates the encoding of classical term rewrite systems. We define the addition on naturals as well as the function $double$ with their usual meanings.

\noindent
$r:+(n:0, m:\bullet) \to r \gg m$\\
$r:+(n:succ(p:\bullet), m:\bullet) \to q:succ(k:+(p,m)) ; r \gg q$\\
$r:double(n:\bullet) \to q:+(n,n) ; r \gg q$\\

\end{exam}

\section{Modal logic}
It is now time to define the syntax and the semantics of the logic of graph modifiers that will be used as a tool to talk about rooted termgraphs.
\subsection{Syntax}
Like the language of propositional dynamic logic, the language of the logic of graph modifiers
is based on the idea of associating with each action $\alpha$ of an action language a modal connective $\lbrack\alpha\rbrack$.
The formula $\lbrack\alpha\rbrack\phi$ is read ``after every terminating execution of $\alpha$, $\phi$ is true''.
Consider, as in section~\ref{sec:graph-rew}, a countable set ${\mathcal F}$ (with typical members denoted $a$, $b$, etc) of edge labels
and a countable set $\Omega$ (with typical members denoted $\omega$, $\pi$, etc) of node labels. These labels are formulas  defined below. A node labeled by $\pi$ is called a \emph{$\pi$ node}.

Formally we define the set of all actions (with typical members denoted $\alpha$, $\beta$, etc)
and the set of all formulas (with typical members denoted $\phi$, $\psi$, etc) as follows:
\begin{itemize}
\item $\alpha$ $::=$ $a\mid U\mid n\mid\vec{n}\mid\phi?\mid(\omega:=_{g}\phi)\mid(\omega:=_{l}\phi)\mid(a+(\phi,\psi))\mid(a-(\phi,\psi))\mid(\alpha;\beta)\mid(\alpha\cup\beta)\mid\alpha^{\star}$,
\item $\phi$ $::$ $\omega\mid\bot\mid\neg\phi\mid(\phi\vee\psi)\mid\lbrack\alpha\rbrack\phi$.
\end{itemize}
We adopt the standard abbreviations for the other Boolean connectives.
Moreover, for all actions $\alpha$ and for all formulas $\phi$, let $\langle\alpha\rangle\phi$ be $\neg\lbrack\alpha\rbrack\neg\phi$.
As usual, we follow the standard rules for omission of the parentheses.
An atomic action is either
an edge label $a$ in ${\mathcal F}$,
the universal action $U$,
a test $\phi?$ or
an update action $n$, $\vec{n}$, $\omega:=_{g}\phi$, $\omega:=_{l}\phi$, $a+(\phi,\psi)$ or $a-(\phi,\psi)$.
$U$ reads ``go anywhere'',
$n$ reads ``add some new node'',
$\vec{n}$ reads ``add some new node and go there'',
$\omega:=_{g}\phi$ reads ``assign to $\omega$ nodes the truth value of $\phi$ everywhere (globally)'',
$\omega:=_{l}\phi$ reads ``assign to $\omega$ the truth value of $\phi$ here (locally)'',
$a+(\phi,\psi)$ reads ``add $a$ edges from all $\phi$ nodes to all $\psi$ nodes'', and
$a-(\phi,\psi)$ reads ``delete$a$ edges from all $\phi$ nodes to all $\psi$ nodes''.
Complex actions are built by means of the regular operators ``$;$'', ``$\cup$'' and ``$^{\star}$''.
An update action is an action without edge labels and without $U$.
An update action is $:=_{l}$-free if no local assignment $\omega:=_{l}\phi$ occurs in it.
\subsection{Semantics}
\newcommand {\power} {\mathcal P}

Like the truth-conditions of the formulas of ordinary modal logics, the truth-conditions of the formulas of the logic of graph modifiers is based on the idea of interpreting, within a rooted termgraph $G$ $=$ $({\mathcal N},{\mathcal E},{\mathcal L}^{n},{\mathcal L}^{e},{\mathcal S},{\mathcal T},n_{0})$, edge labels in ${\mathcal F}$ by sets of edges and node labels in $\Omega$ by sets of nodes. In this section, we consider a more general notion of node labeling functions $\nlabel$ of termgraphs such that nodes can have several labels (propositions). In this case the labeling function has the following profile $\nlabel : \nodes \to \power(\signature)$. Node labeling functions considered in section~\ref{sec:graph-rew} where a node can have at most one label is obviously a particular case.
Let $I_{G}$ be the interpretation function in $G$ of labels defined as follows:
\begin{itemize}
\item $I_{G}(a)$ $=$ $\{e$ $\in$ ${\mathcal E}$: ${\mathcal L}^{e}(e)$ $=$ $a\}$,
\item $I_{G}(\omega)$ $=$ $\{n$ $\in$ ${\mathcal N}$: $\omega \in {\mathcal L}^{n}(n) \}$.
\end{itemize}
For all abstract actions $a$, let $R_{G}(a)$ $=$ $\{(n_{1},n_{2})$: there exists an edge $e$ $\in$ $I_{G}(a)$ such that ${\mathcal S}(e)$ $=$ $n_{1}$ and ${\mathcal T}(e)$ $=$ $n_{2}\}$ be the binary relation interpreting the abstract action $a$ in $G$.
The truth-conditions of the formulas of the logic of graph modifiers are defined by induction as follows:
\begin{itemize}
\item $G$ $\models$ $\omega$ iff $n_{0}$ $\in$ $I_{G}(\omega)$,
\item $G$ $\not\models$ $\bot$,
\item $G$ $\models$ $\neg\phi$ iff $G$ $\not\models$ $\phi$,
\item $G$ $\models$ $\phi\vee\psi$ iff $G$ $\models$ $\phi$ or $G$ $\models$ $\psi$,
\item $G$ $\models$ $\lbrack\alpha\rbrack\phi$ iff for all rooted termgraphs $G^{\prime}$ $=$ $({\mathcal N}^{\prime},{\mathcal E}^{\prime},{{\mathcal L}^{n}}^{\prime},{{\mathcal L}^{e}}^{\prime},{\mathcal S}^{\prime},{\mathcal T}^{\prime},n_{0}^{\prime})$, if $G$ $\longrightarrow_{\alpha}$ $G^{\prime}$ then $G^{\prime}$ $\models$ $\phi$
\end{itemize}
where the binary relations $\longrightarrow_{\alpha}$ are defined by induction as follows:
\begin{itemize}
\item $G$ $\longrightarrow_{a}$ $G^{\prime}$ iff ${\mathcal N}^{\prime}$ $=$ ${\mathcal N}$, ${\mathcal E}^{\prime}$ $=$ ${\mathcal E}$, ${{\mathcal L}^{n}}^{\prime}$ $=$ ${\mathcal L}^{n}$, ${{\mathcal L}^{e}}^{\prime}$ $=$ ${\mathcal L}^{e}$, ${\mathcal S}^{\prime}$ $=$ ${\mathcal S}$, ${\mathcal T}^{\prime}$ $=$ ${\mathcal T}$ and
$(n_{0},n_{0}^{\prime}) \in R_{G}(a)$,
%
%
\item $G$ $\longrightarrow_{\phi?}$ $G^{\prime}$ iff ${\mathcal N}^{\prime}$ $=$ ${\mathcal N}$, ${\mathcal E}^{\prime}$ $=$ ${\mathcal E}$, ${{\mathcal L}^{n}}^{\prime}$ $=$ ${\mathcal L}^{n}$, ${{\mathcal L}^{e}}^{\prime}$ $=$ ${\mathcal L}^{e}$, ${\mathcal S}^{\prime}$ $=$ ${\mathcal S}$, ${\mathcal T}^{\prime}$ $=$ ${\mathcal T}$, $n_{0}^{\prime}$ $=$ $n_{0}$ and $G^{\prime}$ $\models$ $\phi$,
\item $G$ $\longrightarrow_{U}$ $G^{\prime}$ iff ${\mathcal N}^{\prime}$ $=$ ${\mathcal N}$, ${\mathcal E}^{\prime}$ $=$ ${\mathcal E}$, ${{\mathcal L}^{n}}^{\prime}$ $=$ ${\mathcal L}^{n}$, ${{\mathcal L}^{e}}^{\prime}$ $=$ ${\mathcal L}^{e}$, ${\mathcal S}^{\prime}$ $=$ ${\mathcal S}$ and ${\mathcal T}^{\prime}$ $=$ ${\mathcal T}$,
\item $G$ $\longrightarrow_{n}$ $G^{\prime}$ iff ${\mathcal N}^{\prime}$ $=$ ${\mathcal N}\cup\{n_{1}\}$ where $n_{1}$ is a new node, ${\mathcal E}^{\prime}$ $=$ ${\mathcal E}$, ${{\mathcal L}^{n}}^{\prime} (m)$ $=$ ${\mathcal L}^{n} (m)$ if $m \not= n_1$, ${{\mathcal L}^{n}}^{\prime}(n_1) = \emptyset$, ${{\mathcal L}^{e}}^{\prime}$ $=$ ${\mathcal L}^{e}$, ${\mathcal S}^{\prime}$ $=$ ${\mathcal S}$, ${\mathcal T}^{\prime}$ $=$ ${\mathcal T}$ and $n_{0}^{\prime}$ $=$ $n_{0}$,
\item $G$ $\longrightarrow_{\vec{n}}$ $G^{\prime}$ iff ${\mathcal N}^{\prime}$ $=$ ${\mathcal N}\cup\{n_{1}\}$ where $n_{1}$ is a new node, ${\mathcal E}^{\prime}$ $=$ ${\mathcal E}$, ${{\mathcal L}^{n}}^{\prime} (m)$ $=$ ${\mathcal L}^{n}(m)$ if $m \not= n_1$, ${{\mathcal L}^{n}}^{\prime}(n_1) = \emptyset$, ${{\mathcal L}^{e}}^{\prime}$ $=$ ${\mathcal L}^{e}$, ${\mathcal S}^{\prime}$ $=$ ${\mathcal S}$, ${\mathcal T}^{\prime}$ $=$ ${\mathcal T}$ and $n_{0}^{\prime}$ $=$ $n_{1}$,
\item $G$ $\longrightarrow_{\omega:=_{g}\phi}$ $G^{\prime}$ iff ${\mathcal N}^{\prime}$ $=$ ${\mathcal N}$, ${\mathcal E}^{\prime}$ $=$ ${\mathcal E}$, ${{\mathcal L}^{n}}^{\prime}(m)$ $=$ if $({\mathcal N},{\mathcal E},{\mathcal L}^{n},{\mathcal L}^{e},{\mathcal S},{\mathcal T},m)$ $\models$ $\phi\}$ then $\nlabel(m) \cup \{ \omega\}$ else $\nlabel(m) \setminus \{\omega\}$, ${{\mathcal L}^{e}}^{\prime}$ $=$ ${\mathcal L}^{e}$, ${\mathcal S}^{\prime}$ $=$ ${\mathcal S}$, ${\mathcal T}^{\prime}$ $=$ ${\mathcal T}$ and $n_{0}^{\prime}$ $=$ $n_{0}$,

\item $G$ $\longrightarrow_{\omega:=_{l}\phi}$ $G^{\prime}$ iff ${\mathcal N}^{\prime}$ $=$ ${\mathcal N}$, ${\mathcal E}^{\prime}$ $=$ ${\mathcal E}$, ${{\mathcal L}^{n}}^{\prime}(n_0)$ $=$ if $({\mathcal N},{\mathcal E},{\mathcal L}^{n},{\mathcal L}^{e},{\mathcal S},{\mathcal T},n_0)$ $\models$ $\phi$ then ${\mathcal L}^{n}(n_0)\cup\{\omega\}$ else ${\mathcal L}^{n}(n_0)\setminus\{\omega\}$, ${{\mathcal L}^{n}}^{\prime} (m)$ $=$ ${\mathcal L}^{n} (m)$ if $m \not= n_0$, ${{\mathcal L}^{e}}^{\prime}$ $=$ ${\mathcal L}^{e}$, ${\mathcal S}^{\prime}$ $=$ ${\mathcal S}$, ${\mathcal T}^{\prime}$ $=$ ${\mathcal T}$ and $n_{0}^{\prime}$ $=$ $n_{0}$,
\item $G$ $\longrightarrow_{a+(\phi,\psi)}$ $G^{\prime}$ iff ${\mathcal N}^{\prime}$ $=$ ${\mathcal N}$, ${\mathcal E}^{\prime}$ $=$ ${\mathcal E} \cup \{(n_1,a,n_2) : ({\mathcal N},{\mathcal E},{\mathcal L}^{n},{\mathcal L}^{e},{\mathcal S},{\mathcal T},n_{1}) \models \phi$ and $({\mathcal N},{\mathcal E},{\mathcal L}^{n},{\mathcal L}^{e},{\mathcal S},{\mathcal T},n_{2})$ $\models$ $\psi \}$,
${{\mathcal L}^{n}}^{\prime}$ $=$ ${\mathcal L}^{n}$, ${{\mathcal
L}^{e}}^{\prime}(e)$ $=$ if $e \in {\mathcal E}$ then ${\mathcal
L}^{e}(e)$ else $a$, ${\mathcal S}^{\prime} (e)$ $=$ if $e \in
{\mathcal E}$ then ${\mathcal S}(e)$ else $e$ is of the form $(n_1, a,
n_2)$ and ${\mathcal S}^{\prime} (e) = n_1$, ${\mathcal T}^{\prime}$
$=$ if $e \in {\mathcal E}$ then ${\mathcal T}(e)$ else $e$ is of the
form $(n_1, a, n_2)$ and ${\mathcal T}^{\prime} (e) = n_2$ and
$n_{0}^{\prime}$ $=$ $n_{0}$,
\item $G$ $\longrightarrow_{a-(\phi,\psi)}$ $G^{\prime}$ iff ${\mathcal N}^{\prime}$ $=$ ${\mathcal N}$, ${\mathcal E}^{\prime}$ $=$ ${\mathcal E} \setminus \{(n_1,a,n_2) : ({\mathcal N},{\mathcal E},{\mathcal L}^{n},{\mathcal L}^{e},{\mathcal S},{\mathcal T},n_{1}) \models \phi$ and $({\mathcal N},{\mathcal E},{\mathcal L}^{n},{\mathcal L}^{e},{\mathcal S},{\mathcal T},n_{2})$ $\models$ $\psi \}$, ${{\mathcal L}^{n}}^{\prime}$ $=$ ${\mathcal L}^{n}$, ${{\mathcal L}^{e}}^{\prime}(e)$ $=$ ${\mathcal L}^{e}(e)$, ${\mathcal S}^{\prime}$ $=$ ${\mathcal S}$, ${\mathcal T}^{\prime}$ $=$ ${\mathcal T}$ and $n_{0}^{\prime}$ $=$ $n_{0}$,
\item $G$ $\longrightarrow_{\alpha;\beta}$ $G^{\prime}$ iff there exists a rooted termgraph $G^{\prime\prime}$ $=$ $({\mathcal N}^{\prime\prime},{\mathcal E}^{\prime\prime},{{\mathcal L}^{n}}^{\prime\prime},{{\mathcal L}^{e}}^{\prime\prime},{\mathcal S}^{\prime\prime},{\mathcal T}^{\prime\prime},n_{0}^{\prime\prime})$ such that $G$ $\longrightarrow_{\alpha}$ $G^{\prime\prime}$ and $G^{\prime\prime}$ $\longrightarrow_{\beta}$ $G^{\prime}$,
\item $G$ $\longrightarrow_{\alpha\cup\beta}$ $G^{\prime}$ iff $G$ $\longrightarrow_{\alpha}$ $G^{\prime}$ or $G$ $\longrightarrow_{\beta}$ $G^{\prime}$,
\item $G$ $\longrightarrow_{\alpha^{\star}}$ $G^{\prime}$ iff there exists a sequence $G^{(0)}$ $=$ $({\mathcal N}^{(0)},{\mathcal E}^{(0)},{{\mathcal L}^{n}}^{(0)},{{\mathcal L}^{e}}^{(0)},{\mathcal S}^{(0)},{\mathcal T}^{(0)},n_{0}^{(0)})$, $\ldots$, $G^{(k)}$ $=$ $({\mathcal N}^{(k)},{\mathcal E}^{(k)},{{\mathcal L}^{n}}^{(k)},{{\mathcal L}^{e}}^{(k)},{\mathcal S}^{(k)},{\mathcal T}^{(k)},n_{0}^{(k)})$ of rooted termgraphs such that $G^{(0)}$ $=$ $G$, $G^{(k)}$ $=$ $G^{\prime}$ and for all non-negative integers $i$, if $i$ $<$ $k$ then $G^{(i)}$ $\longrightarrow_{\alpha}$ $G^{(i+1)}$.
\end{itemize}
The above definitions of formulas  reflect our intuitive understanding of the actions of the language of the logic of graph modifiers.
Obviously, $G$ $\models$ $\langle\alpha\rangle\phi$ iff there exists a rooted termgraph $G^{\prime}$ $=$ $({\mathcal N}^{\prime},{\mathcal E}^{\prime},{{\mathcal L}^{n}}^{\prime},{{\mathcal L}^{e}}^{\prime},{\mathcal S}^{\prime},{\mathcal T}^{\prime},n_{0}^{\prime})$ such that $G$ $\longrightarrow_{\alpha}$ $G^{\prime}$ and $G^{\prime}$ $\models$ $\phi$.
The formula $\phi$ is said to be valid in class ${\mathcal C}$ of rooted termgraphs, in symbols ${\mathcal C}$ $\models$ $\phi$, iff $G$ $\models$ $\phi$ for each rooted termgraph $G$ $=$ $({\mathcal N},{\mathcal E},{\mathcal L}^{n},{\mathcal L}^{e},{\mathcal S},{\mathcal T},n_{0})$ in ${\mathcal C}$.
The class of all rooted termgraphs will be denoted more briefly as ${\mathcal C}_{all}$.
\subsection{Validities}
Obviously, as in propositional dynamic logic, we have
\begin{itemize}
\item ${\mathcal C}_{all}$ $\models$ $\lbrack\phi?\rbrack\psi\leftrightarrow(\phi\rightarrow\psi)$,
\item ${\mathcal C}_{all}$ $\models$ $\lbrack\alpha;\beta\rbrack\phi\leftrightarrow\lbrack\alpha\rbrack\lbrack\beta\rbrack\phi$,
\item ${\mathcal C}_{all}$ $\models$ $\lbrack\alpha\cup\beta\rbrack\phi\leftrightarrow\lbrack\alpha\rbrack\phi\wedge\lbrack\beta\rbrack\phi$,
\item ${\mathcal C}_{all}$ $\models$ $\lbrack\alpha^{\star}\rbrack\phi\leftrightarrow\phi\wedge\lbrack\alpha\rbrack\lbrack\alpha^{\star}\rbrack\phi$.
\end{itemize}
If $\alpha$ is a $:=_{l}$-free update action then
\begin{itemize}
\item ${\mathcal C}_{all}$ $\models$ $\lbrack\alpha\rbrack\bot\leftrightarrow\bot$,
\item ${\mathcal C}_{all}$ $\models$ $\lbrack\alpha\rbrack\neg\phi\leftrightarrow\neg\lbrack\alpha\rbrack\phi$,
\item ${\mathcal C}_{all}$ $\models$ $\lbrack\alpha\rbrack(\phi\vee\psi)\leftrightarrow\lbrack\alpha\rbrack\phi\vee\lbrack\alpha\rbrack\psi$.
\end{itemize}
The next series of equivalences guarantees that each of our $:=_{l}$-free update actions can be moved across the abstract actions of the form $a$ or $U$:
\begin{itemize}
\item ${\mathcal C}_{all}$ $\models$ $\lbrack n\rbrack\lbrack a\rbrack\phi\leftrightarrow\lbrack a\rbrack\lbrack n\rbrack\phi$,
\item ${\mathcal C}_{all}$ $\models$ $\lbrack n\rbrack\lbrack U\rbrack\phi\leftrightarrow\lbrack n\rbrack\phi\wedge\lbrack U\rbrack\lbrack n\rbrack\phi$,
\item ${\mathcal C}_{all}$ $\models$ $\lbrack\vec{n}\rbrack\lbrack a\rbrack\phi\leftrightarrow\top$,
\item ${\mathcal C}_{all}$ $\models$ $\lbrack\vec{n}\rbrack\lbrack U\rbrack\phi\leftrightarrow\lbrack\vec{n}\rbrack\phi\wedge\lbrack U\rbrack\lbrack n\rbrack\phi$,
\item ${\mathcal C}_{all}$ $\models$ $\lbrack\omega:=_{g}\phi\rbrack\lbrack a\rbrack\psi\leftrightarrow\lbrack a\rbrack\lbrack\omega:=_{g}\phi\rbrack\phi$,
\item ${\mathcal C}_{all}$ $\models$ $\lbrack\omega:=_{g}\phi\rbrack\lbrack U\rbrack\psi\leftrightarrow\lbrack U\rbrack\lbrack\omega:=_{g}\phi\rbrack\psi$,
\item ${\mathcal C}_{all}$ $\models$ $\lbrack a+(\phi,\psi)\rbrack\lbrack b\rbrack\chi\leftrightarrow\lbrack b\rbrack\lbrack a+(\phi,\psi)\rbrack\chi$ if $a$ $\not=$ $b$ and ${\mathcal C}_{all}$ $\models$ $\lbrack a+(\phi,\psi)\rbrack\lbrack b\rbrack\chi\leftrightarrow\lbrack b\rbrack\lbrack a+(\phi,\psi)\rbrack\chi\wedge(\phi\rightarrow\lbrack U\rbrack(\psi\rightarrow\lbrack a+(\phi,\psi)\rbrack\chi))$ if $a$ $=$ $b$,
\item ${\mathcal C}_{all}$ $\models$ $\lbrack a+(\phi,\psi)\rbrack\lbrack U\rbrack\chi\leftrightarrow\lbrack U\rbrack\lbrack a+(\phi,\psi)\rbrack\chi$,
\item ${\mathcal C}_{all}$ $\models$ $\lbrack a-(\phi,\psi)\rbrack\lbrack b\rbrack\chi\leftrightarrow\lbrack b\rbrack\lbrack a-(\phi,\psi)\rbrack\chi$ if $a$ $\not=$ $b$ and ${\mathcal C}_{all}$ $\models$ $\lbrack a-(\phi,\psi)\rbrack\lbrack b\rbrack\chi\leftrightarrow(\neg\phi\wedge\lbrack b\rbrack\lbrack a-(\phi,\psi)\rbrack\chi)\vee(\phi\wedge\lbrack b\rbrack(\neg\psi\rightarrow\lbrack a-(\phi,\psi)\rbrack\chi))$ if $a$ $=$ $b$,
\item ${\mathcal C}_{all}$ $\models$ $\lbrack a-(\phi,\psi)\rbrack\lbrack U\rbrack\chi\leftrightarrow\lbrack U\rbrack\lbrack a-(\phi,\psi)\rbrack\chi$.
\end{itemize}
Finally, once we have moved each of our $:=_{l}$-free update actions across the abstract actions of the form $a$ or $U$, these update actions can be eliminated by means of the following equivalences:
\begin{itemize}
\item ${\mathcal C}_{all}$ $\models$ $\lbrack n\rbrack\omega\leftrightarrow\omega$,
\item ${\mathcal C}_{all}$ $\models$ $\lbrack\vec{n}\rbrack\omega\leftrightarrow\bot$,
\item ${\mathcal C}_{all}$ $\models$ $\lbrack\omega:=_{g}\phi\rbrack\pi\leftrightarrow\pi$ if $\omega$ $\not=$ $\pi$ and ${\mathcal C}_{all}$ $\models$ $\lbrack\omega:=_{g}\phi\rbrack\pi\leftrightarrow\phi$ if $\omega$ $=$ $\pi$,
\item ${\mathcal C}_{all}$ $\models$ $\lbrack a+(\phi,\psi)\rbrack\omega\leftrightarrow\omega$,
\item ${\mathcal C}_{all}$ $\models$ $\lbrack a-(\phi,\psi)\rbrack\omega\leftrightarrow\omega$.
\end{itemize}
\begin{prop}\label{pro_1}
For all $:=_{l}$-free $^{\star}$-free formulas $\phi$, there exists a $:=_{l}$-free $^{\star}$-free formula $\psi$ without update actions such that ${\mathcal C}_{all}$ $\models$ $\phi\leftrightarrow\psi$.
\end{prop}
\begin{proof}
See the above discussion.
\end{proof}
\ \\

Just as for $:=_{l}$-free update actions, we have the following equivalences for the update actions of the form $\omega:=_{l}\phi$:
\begin{itemize}
\item ${\mathcal C}_{all}$ $\models$ $\lbrack\omega:=_{l}\phi\rbrack\bot\leftrightarrow\bot$,
\item ${\mathcal C}_{all}$ $\models$ $\lbrack\omega:=_{l}\phi\rbrack\neg\psi\leftrightarrow\neg\lbrack\omega:=_{l}\phi\rbrack\psi$,
\item ${\mathcal C}_{all}$ $\models$ $\lbrack\omega:=_{l}\phi\rbrack(\psi\vee\chi)\leftrightarrow\lbrack\omega:=_{l}\phi\rbrack\psi\vee\lbrack\omega:=_{l}\phi\rbrack\chi$,
\item ${\mathcal C}_{all}$ $\models$ $\lbrack\omega:=_{l}\phi\rbrack\pi\leftrightarrow\pi$ if $\omega$ $\not=$ $\pi$ and ${\mathcal C}_{all}$ $\models$ $\lbrack\omega:=_{g}\phi\rbrack\pi\leftrightarrow\phi$ if $\omega$ $=$ $\pi$.
\end{itemize}
But it is not possible to formulate reduction axioms for the cases $\lbrack\omega:=_{l}\phi\rbrack\lbrack a\rbrack\psi$ and $\lbrack\omega:=_{l}\phi\rbrack\lbrack U\rbrack\psi$.
More precisely,
\begin{prop}\label{pro_2}
There exists a $^{\star}$-free formula $\phi$ such that for all $^{\star}$-free formulas $\psi$ without update actions, ${\mathcal C}_{all}$ $\not\models$ $\phi\leftrightarrow\psi$.
\end{prop}
\begin{proof}
Take the $^{\star}$-free formula $\phi$ $=$ $\lbrack\omega:=_{g}\bot\rbrack\lbrack U\rbrack\lbrack\omega:=_{l}\top\rbrack\lbrack a\rbrack\neg\omega$.
The reader may easily verify that for all rooted termgraphs $G$ $=$ $({\mathcal N},{\mathcal E},{\mathcal L}^{n},{\mathcal L}^{e},{\mathcal S},{\mathcal T},n_{0})$, $G$ $\models$ $\phi$ iff $R_{G}(a)$ is irreflexive.
Seeing that the fact that the binary relation interpreting an abstract action of the form $a$ is irreflexive cannot be modally defined in propositional dynamic logic, then for all formulas $\psi$ without update actions, ${\mathcal C}_{all}$ $\not\models$ $\phi\leftrightarrow\psi$.
\end{proof}
\subsection{Decidability, axiomatization and a link with hybrid logics}
Firstly, let us consider the set $L$ of all $:=_{l}$-free $^{\star}$-free formulas $\phi$ such that ${\mathcal C}_{all}$ $\models$ $\phi$.
Together with a procedure for deciding membership in $^{\star}$-free propositional dynamic logic,
the equivalences preceding proposition~\ref{pro_1} provide a procedure for deciding membership in $L$.
Hence, membership in $L$ is decidable.

Secondly, let us consider the set $L(:=_{l})$ of all $^{\star}$-free formulas $\phi$ such that ${\mathcal C}_{all}$ $\models$ $\phi$.
Aucher {\it et al.}~\cite{AucherBalbianiFarinasHerzig-Entcs09} 
have defined a recursive translation
from the language of hybrid logic~\cite{ArecesTenCate06} into the set of all our $^{\star}$-free formulas that preserves satisfiability.
It is known that the problem of deciding satisfiability of hybrid logic formulas is undecidable
\cite[Section 4.4]{ArecesBlackburnMarx99}.
The language of hybrid logic has formulas of the form
$ @_i \phi$ (``$\phi$ is true at $i$''),
$@_x \phi$ (``$\phi$ is true at $x$'') and
$\ \downarrow \!\! x.\phi$ (``$\phi$ holds after $x$ is bound to the current state''), where
$\mathit{NOM}  = \{i_1, \ldots\}$ is a set of nominals, and
$\mathit{SVAR} = \{x_1, \ldots\}$ is a set of state variables.
The (slightly adapted)
translation of a given hybrid formula $\phi_0 $ is recursively defined as follows.
\begin{displaymath}\begin{array}{lll}
   \tau(\omega)            &=& \omega
\\ \tau( i )             &=& \omega_i        \mbox{ \ \ where } \omega_i \mbox{ does not occur in } \phi_0
\\ \tau( x  )             &=& \omega_x        \mbox{ \ \ where } \omega_x \mbox{ does not occur in } \phi_0
\\ \tau(\lnot \phi)       &=& \lnot \tau(\phi)
\\ \tau(\phi\lor \psi)    &=& \tau(\phi) \lor \tau(\psi)
\\ \tau(\lbrack a \rbrack  \phi)    &=& \lbrack a \rbrack \tau(\phi)
\\ \tau(\lbrack U \rbrack\phi)    &=& \lbrack U \rbrack \tau(\phi)
\\ \tau(@_i    \phi)      &=& \langle U \rangle(\omega_i \land \tau(\phi))
\\ \tau(@_x    \phi)      &=& \langle U \rangle(\omega_x \land \tau(\phi))
\\ \tau(\!\!\ \downarrow \!\! x.\phi) &=& \lbrack{\omega_x :=_g \bot}\rbrack \lbrack{\omega_x :=_l \top}\rbrack \tau(\phi)
\end{array}\end{displaymath}
As the satisfiability problem is undecidable in hybrid logic, membership in $L(:=_{l})$ is undecidable, too.

Thirdly, let us consider the set $L(^{\star})$ of all $:=_{l}$-free formulas $\phi$ such that ${\mathcal C}_{all}$ $\models$ $\phi$.
It is still an open problem whether membership in $L(^{\star})$ is decidable or not:
while the update actions can be eliminated from $:=_{l}$-free formulas, it is not clear whether
this can be done for formulas in which e.g.\ iterations of assignments occur.

As for the axiomatization issue, the equivalences preceding proposition~\ref{pro_1} provide
a sound and complete axiom system of $L$,
whereas no axiom system of $L(:=_{l})$ and $L(^{\star})$ is known to be sound and complete.
\section{Definability of classes of termgraphs}
\label{sec-definability}
For all abstract actions $a$, by means of the update actions of the form $\omega:=_{l}\phi$, we can express the fact that the binary relation interpreting an abstract action of the form $a$ is deterministic, irreflexive or locally reflexive.
More precisely, for all rooted termgraphs $G$ $=$ $({\mathcal N},{\mathcal E},{\mathcal L}^{n},{\mathcal L}^{e},{\mathcal S},{\mathcal T}, n_0)$,
\begin{itemize}
\item $G$ $\models$ $\lbrack\omega:=_{g}\bot\rbrack\lbrack\pi:=_{g}\bot\rbrack\lbrack U\rbrack\lbrack\omega:=_{l}\top\rbrack\lbrack a\rbrack\lbrack\pi:=_{l}\top\rbrack\lbrack U\rbrack(\omega\rightarrow\lbrack a\rbrack\pi)$ iff $R_{G}(a)$ is deterministic,
\item $G$ $\models$ $\lbrack\omega:=_{g}\bot\rbrack\lbrack U\rbrack\lbrack\omega:=_{l}\top\rbrack\lbrack a\rbrack\neg\omega$ iff $R_{G}(a)$ is irreflexive,
\item $G$ $\models$ $\lbrack\omega:=_{g}\bot\rbrack\lbrack\omega:=_{l}\top\rbrack\langle a\rangle\omega$ iff $R_{G}(a)$ is locally reflexive in $n_{0}$.
\end{itemize}

Together with the update actions of the form $\omega:=_{l}\phi$, the regular operation ``$^{\star}$'' enables us to define non-elementary classes of rooted termgraphs.
As a first example, the class of all infinite rooted termgraphs cannot be modally defined in propositional dynamic logic but the following formula pins it down:
\begin{itemize}
\item $\lbrack\omega:=_{g}\top\rbrack\lbrack(U;\omega?;\omega:=_{l}\bot)^{\star}\rbrack\langle U\rangle\omega$.
\end{itemize}
As a second example, take the class of all $a$-cycle-free rooted termgraphs.
It cannot be modally defined in propositional dynamic logic but the following formula pins it down:
\begin{itemize}
\item $\lbrack\omega:=_{g}\top\rbrack\lbrack U\rbrack\lbrack\omega:=_{l}\bot\rbrack\lbrack a^{+}\rbrack\omega$.
\end{itemize}
As a third example, within the class of all $a$-deterministic rooted termgraphs,
the class of all $a$-circular rooted termgraphs\footnote{
In an $a$-circular rooted termgraph for every node $n$ there is an $i$ and there are
$a_1$, \ldots $a_n$ such that $a = a_1 = a_n$ and  $n_k$ is related to $n_{k+1}$ by an edge labelled $a$,
for all $k\leq i$.
}
cannot be modally defined in propositional dynamic logic but the following formula pins it down:
\begin{itemize}
\item $\lbrack\omega:=_{g}\bot\rbrack\lbrack U\rbrack\lbrack\omega:=_{l}\top\rbrack\langle a^{+}\rangle\omega$.
\end{itemize}
Now, within the class of all rooted termgraphs that are both $a$- and $b$-deterministic,
the class of all $(a\leq b)$ rooted termgraphs
\footnote{
Rooted termgraphs are termgraphs where the path obtained by following feature $b$
is longer than or equal to the path obtained by following feature $a$.}
cannot be modally defined in propositional dynamic logic but the following formula pins it down:
\begin{itemize}
\item $\lbrack\omega:=_{g}\bot\rbrack\lbrack\omega:=_{l}\top\rbrack\lbrack\pi:=_{g}\bot\rbrack\lbrack\pi:=_{l}\top\rbrack\lbrack((U;\omega?;a;\neg\omega?;\omega:=_{l}\top);(U;\pi?;b;\neg\pi?;\pi:=_{l}\top))^{\star}\rbrack(\langle U\rangle(\pi\wedge\lbrack b\rbrack\bot)\rightarrow\langle U\rangle(\omega\wedge\lbrack a\rbrack\bot))$.
\end{itemize}
Finally, within the class of all finite $(a\cup b)$-cycle-free $(a,b)$-deterministic rooted termgraphs, the class of all $(a,b)$-binary rooted termgraphs cannot be modally defined in propositional dynamic logic but the following formula pins it down:
\begin{itemize}
\item $\lbrack\omega:=_{g}\bot\rbrack\lbrack U\rbrack\lbrack\omega:=_{l}\top\rbrack\lbrack a\rbrack\lbrack\pi:=_{g}\top\rbrack\lbrack(a\cup b)^{\star}\rbrack\lbrack\pi:=_{l}\bot\rbrack\lbrack U\rbrack(\omega\rightarrow\lbrack b\rbrack\lbrack(a\cup b)^{\star}\rbrack\pi)$.
\end{itemize}
Most important of all is the ability of the language of the logic of graph modifiers to characterize finite graph homomorphisms.
\begin{prop}\label{pro_3}
Let $G$ $=$ $({\mathcal N},{\mathcal E},{\mathcal L}^{n},{\mathcal L}^{e},{\mathcal S},{\mathcal T},n_{0})$ be a finite rooted termgraph.
There exists a $^{\star}$-free action $\alpha_{G}$ and a $^{\star}$-free formula $\phi_{G}$ such that for all finite rooted termgraphs $G^{\prime}$ $=$ $({\mathcal N}^{\prime},{\mathcal E}^{\prime},{{\mathcal L}^{n}}^{\prime},{{\mathcal L}^{e}}^{\prime},{\mathcal S}^{\prime},{\mathcal T}^{\prime},n_{0}^{\prime})$, $G^{\prime}$ $\models$ $\langle\alpha_{G}\rangle\phi_{G}$ iff there exists a graph homomorphism from $G$ into $G^{\prime}$.
\end{prop}
\begin{proof}
Let $G$ $=$ $({\mathcal N},{\mathcal E},{\mathcal L}^{n},{\mathcal L}^{e},{\mathcal S},{\mathcal T},n_{0})$ be a finite rooted termgraph.
Suppose that ${\mathcal N}$ $=$ $\{0,\ldots,N-1\}$ and consider a sequence $(\pi_{0},\ldots,\pi_{N-1})$ of pairwise distinct elements of $\Omega$.
Each $\pi_i$ will identify exactly one node of ${\mathcal N}$, and $\pi_{0}$ will identify the root.

We define the action $\alpha_{G}$ and the formula $\phi_{G}$ as follows:
\begin{itemize}
\item $\beta_{G}$ $=$ $(\pi_{0}:=_{g}\bot);\ldots;(\pi_{N-1}:=_{g}\bot)$,
\item for all non-negative integers $i$, if $i$ $<$ $N$ then
$\gamma_{G}^{i}$ $=$ $(\neg\pi_{0}\wedge\ldots\wedge\neg\pi_{i-1})?;(\pi_{i}:=_{l}\top);U$,
\item $\alpha_{G}$ $=$ $\beta_{G};\gamma_{G}^{0};\ldots;\gamma_{G}^{N-1}$,
\item for all non-negative integers $i$, if $i$ $<$ $N$ then $\psi_{G}^{i}$ $=$ if ${\mathcal L}^{n}(i)$ is defined then $\langle U\rangle(\pi_{i}\wedge{\mathcal L}^{n}(i))$ else $\top$,
\item for all non-negative integers $i,j$, if $i,j$ $<$ $N$ then $\chi_{G}^{i,j}$ $=$ if there exists an edge $e$ $\in$ ${\mathcal E}$ such that ${\mathcal S}(e)$ $=$ $i$ and ${\mathcal T}(e)$ $=$ $j$ then $\langle U\rangle(\pi_{i}\wedge\langle{\mathcal L}^{e}(e)\rangle\pi_{j})$ else $\top$,
\item $\phi_{G}$ $=$ $\psi_{G}^{0}\wedge\ldots\wedge\psi_{G}^{N-1}\wedge\chi_{G}^{0,0}\wedge\ldots\wedge\chi_{G}^{N-1,N-1}$.
\end{itemize}
The reader may easily verify that for all finite rooted termgraphs \\
$G^{\prime}$ $=$ $({\mathcal N}^{\prime},{\mathcal E}^{\prime},{{\mathcal L}^{n}}^{\prime},{{\mathcal L}^{e}}^{\prime},{\mathcal S}^{\prime},{\mathcal T}^{\prime},n_{0}^{\prime})$, $G^{\prime}$ $\models$ $\langle\alpha_{G}\rangle\phi_{G}$ iff there exists a graph homomorphism from $G$ to $G^{\prime}$.
\end{proof}
\section{Definability of transformations of termgraphs }
\label{sec-logic-transformation}
In this section we show how elementary actions over termgraphs as defined in Section~\ref{sec:graph-rew} can be encoded by means of formulas of the proposed modal logic.
Let $\alpha_{a}$ be the action defined as follows:
\begin{itemize}
\item $\alpha_{a}$ $=$ $(\omega:=_{g}\bot);(\omega:=_{l}\top);(\pi:=_{g}\bot);(\pi:=_{g}\langle a\rangle\omega);(a-(\top,\omega));\vec{n};(\omega:=_{g}\bot);(\omega:=_{l}\top);(a+(\pi,\omega))$.
\end{itemize}
The reader may easily verify that for all rooted termgraphs \\
$G$ $=$ $({\mathcal N},{\mathcal E},{{\mathcal L}^{n}},{{\mathcal L}^{e}},{\mathcal S},{\mathcal T},n_{0})$ and
$G^{\prime}$ $=$ $({\mathcal N}^{\prime},{\mathcal E}^{\prime},{{\mathcal L}^{n}}^{\prime},{{\mathcal L}^{e}}^{\prime},{\mathcal S}^{\prime},{\mathcal T}^{\prime},n_{0}^{\prime})$,
$G$ $\longrightarrow_{\alpha_{a}}$ $G^{\prime}$ iff $G^{\prime}$ is obtained from $G$ by redirecting every $a$-edge pointing to the current root towards a freshly created new root.
Hence, together with the update actions $n$, $\vec{n}$, $\omega:=_{g}\phi$, $\omega:=_{l}\phi$, $a+(\phi,\psi)$ and $a-(\phi,\psi)$,
the regular operations ``$;$'', ``$\cup$'' and ``$^{\star}$'' enable us to
define the elementary actions of node labelling, local redirection and global redirection of Section~\ref{sec:graph-rew}.
Let us firstly consider the elementary action of node labelling: $n:f(a_{1}\Rightarrow n_{1},\ldots,a_{k}\Rightarrow n_{k})$.
Applying this elementary action consists in redirecting towards nodes $n_{1}$, $\ldots$, $n_{k}$ the targets of $a_{1}$-, $\ldots$, $a_{k}$- edges starting from node $n$.
It corresponds to the action $nl(n:f(a_{1}\Rightarrow n_{1},\ldots,a_{k}\Rightarrow n_{k}))$ defined as follows:
\begin{itemize}
\item $nl(n:f(a_{1}\Rightarrow n_{1},\ldots,a_{k}\Rightarrow n_{k}))$ $=$ $U;\pi_{n}?;(f:=_{l}\top);(a_{1}+(\pi_{n},\pi_{n_{1}}));\ldots;(a_{k}+(\pi_{n},\pi_{n_{k}}))$.
\end{itemize}
where the $\pi_{i}$'s are as in the proof of Proposition \ref{pro_3}.
The reader may easily verify that for all rooted termgraphs $G$ $=$ $({\mathcal N},{\mathcal E},{{\mathcal L}^{n}},{{\mathcal L}^{e}},{\mathcal S},{\mathcal T},n_{0})$, $G^{\prime}$ $=$ $({\mathcal N}^{\prime},{\mathcal E}^{\prime},{{\mathcal L}^{n}}^{\prime},{{\mathcal L}^{e}}^{\prime},{\mathcal S}^{\prime},{\mathcal T}^{\prime},n_{0}^{\prime})$, $G$ $\longrightarrow_{nl(n:f(a_{1}\Rightarrow n_{1},\ldots,a_{k}\Rightarrow n_{k}))}$ $G^{\prime}$ iff $G^{\prime}$ is obtained from $G$ by redirecting towards nodes $n_{1}$, $\ldots$, $n_{k}$ the targets of $a_{1}$-, $\ldots$, $a_{k}$- edges starting from node $n$.
Let us secondly consider the elementary action of local redirection: $n\gg^{l}_{a}m$.
Applying this elementary action consists in redirecting towards node $m$ the target of an $a$-edge starting from node $n$.
It corresponds to the action $lr(n,a,m)$ defined as follows:
\begin{itemize}
\item $lr(n,a,m)$ $=$ $(a-(\pi_{n},\top));(a+(\pi_{n},\pi_{m}))$.
\end{itemize}
The reader may easily verify that for all rooted termgraphs $G$ $=$ $({\mathcal N},{\mathcal E},{{\mathcal L}^{n}},{{\mathcal L}^{e}},{\mathcal S},{\mathcal T},n_{0})$, $G^{\prime}$ $=$ $({\mathcal N}^{\prime},{\mathcal E}^{\prime},{{\mathcal L}^{n}}^{\prime},{{\mathcal L}^{e}}^{\prime},{\mathcal S}^{\prime},{\mathcal T}^{\prime},n_{0}^{\prime})$, $G$ $\longrightarrow_{lr(n,a,m)}$ $G^{\prime}$ iff $G^{\prime}$ is obtained from $G$ by redirecting towards node $m$ the target of an $a$-edge starting from node $n$.
Let us thirdly consider the elementary action of global redirection: $n\gg^{g}_{a}m$.
Applying this elementary action consists in redirecting towards node $n$ the target of every $a$-edge pointing towards node $m$.
It corresponds to the action $gr(n,a,m)$ defined as follows:
\begin{itemize}
\item $gr(n,a,m)$ $=$ $(\lambda_a:=_{g}\bot);(\lambda_a:=_{g}\langle a\rangle\pi_{n});(a-(\top,\pi_{n}));(a+(\lambda_a,\pi_{m}))$.
\end{itemize}
The reader may easily verify that for all rooted termgraphs $G$ $=$ $({\mathcal N},{\mathcal E},{{\mathcal L}^{n}},{{\mathcal L}^{e}},{\mathcal S},{\mathcal T},n_{0})$, $G^{\prime}$ $=$ $({\mathcal N}^{\prime},{\mathcal E}^{\prime},{{\mathcal L}^{n}}^{\prime},{{\mathcal L}^{e}}^{\prime},{\mathcal S}^{\prime},{\mathcal T}^{\prime},n_{0}^{\prime})$, $G$ $\longrightarrow_{gr(n,a,m)}$ $G^{\prime}$ iff $G^{\prime}$ is obtained from $G$ by redirecting towards node $n$ the target of every $a$-edge pointing towards node $m$. To redirect towards $n$
the target of all edges pointing towards $m$, the action $gr(n,a,m)$ can be performed for all $a \in \features$. We get $gr(n,m) = \bigwedge\limits_{a \in \features}^{} gr(n,a,m)$.
\section{Translating rewrite rules in modal logic}
\label{sec-traduction}

Now we are ready to show how termgraph rewriting can be specified by means of formulas of the proposed  modal logic.

Let $G\rightarrow(a_{1},\ldots,a_{n})$ be a rewrite rule as defined in Section \ref{sec:graph-rew},
i.e., $G$ $=$ $({\mathcal N},{\mathcal E},{{\mathcal L}^{n}},{{\mathcal L}^{e}},{\mathcal S},{\mathcal T},n_{0})$ is a finite rooted termgraph and $(a_{1},\ldots,a_{n})$ is a finite sequence of elementary actions.
We have seen how to associate to $G$ a $^{\star}$-free action $\alpha_{G}$ and a $^{\star}$-free formula $\phi_{G}$ such that for all finite rooted termgraphs $G^{\prime}$ $=$ $({\mathcal N}^{\prime},{\mathcal E}^{\prime},{{\mathcal L}^{n}}^{\prime},{{\mathcal L}^{e}}^{\prime},{\mathcal S}^{\prime},{\mathcal T}^{\prime},n_{0}^{\prime})$, $G^{\prime}$ $\models$ $\langle\alpha_{G}\rangle\phi_{G}$ iff there exists a graph homomorphism from $G$ into $G^{\prime}$.
We have also seen how to associate to the elementary actions $a_{1}$, $\ldots$, $a_{n}$ actions $\alpha_{1}$, $\ldots$, $\alpha_{n}$.
In the following proposition we show how to formulate the fact that a normal form with respect to a rewrite rule (generalization to a set of rules is obvious) satisfies a given formula $\varphi$. A termgraph $t$ is in normal form with respect to a rule $R$ iff $t$ cannot be rewritten by means of $R$. Such formulation may help to express proof obligations of programs specified as termgraph rewrite rules.
Let $n_{1}$, $\ldots$, $n_{k}$ be the list of all nodes occurring in $a_{1}$, $\ldots$, $a_{n}$ but not occurring in $G$.
The truth of the matter is that
\begin{prop}\label{pro_4}
Let $\varphi$ be a modal formula.
For all finite rooted termgraphs $G^{\prime}$ $=$ $({\mathcal N}^{\prime},{\mathcal E}^{\prime},{{\mathcal L}^{n}}^{\prime},{{\mathcal L}^{e}}^{\prime},{\mathcal S}^{\prime},{\mathcal T}^{\prime},n_{0}^{\prime})$, every normal form of $G^{\prime}$ with respect to $G\rightarrow(a_{1},\ldots,a_{n})$ satisfies $\varphi$ iff $G^{\prime}$ $\models$ $\lbrack(\alpha_{G};\phi_{G}?;\vec{n};(\pi_{n_{1}}:=_{g}\bot);(\pi_{n_{1}}:=_{l}\top);\ldots;\vec{n};(\pi_{n_{k}}:=_{g}\bot);(\pi_{n_{k}}:=_{l}\top);\alpha_{1};\ldots;\alpha_{n})^{\star}\rbrack(\lbrack\alpha_{G};\phi_{G}?\rbrack\bot\rightarrow\varphi)$.
\end{prop}
\begin{proof}
$\Leftarrow$: Suppose that $G^{\prime}$ $\models$ $\lbrack(\alpha_{G};\phi_{G}?;\vec{n};(\pi_{n_{1}}:=_{g}\bot);(\pi_{n_{1}}:=_{l}\top);\ldots;\vec{n};(\pi_{n_{k}}:=_{g}\bot);(\pi_{n_{k}}:=_{l}\top);\alpha_{1};\ldots;\alpha_{n})^{\star}\rbrack(\lbrack\alpha_{G};\phi_{G}?\rbrack\bot\rightarrow\varphi)$.
Consider a normal form $G^{nf}$ of $G^{\prime}$ with respect to $G\rightarrow(a_{1},\ldots,a_{n})$.
Then there exists a non-negative integer $k$ and there exist finite rooted termgraphs $G^{0}$, $\ldots$, $G^{k}$ such that:
\begin{itemize}
\item $G^{0}$ $=$ $G^{\prime}$,
\item $G^{k}$ $=$ $G^{nf}$,
\item for all non-negative integers $i$, if $i$ $<$ $k$ then $G_{i}$ $\rightarrow_{G\rightarrow(a_{1},\ldots,a_{n})}$ $G_{i+1}$.
\end{itemize}
Hence, for all non-negative integers $i$, if $i$ $<$ $k$ then \\
$G_{i}$ $\longrightarrow_{\alpha_{G};\phi_{G}?;\vec{n};(\pi_{n_{1}}:=_{g}\bot);(\pi_{n_{1}}:=_{l}\top);\ldots;\vec{n};(\pi_{n_{k}}:=_{g}\bot);(\pi_{n_{k}}:=_{l}\top);\alpha_{1};\ldots;\alpha_{n}}$ $G_{i+1}$.
Moreover, seeing that $G^{nf}$ is a normal form with respect to $G\rightarrow(a_{1},\ldots,a_{n})$, $G^{nf}$ $\models$ $\lbrack\alpha_{G};\phi_{G}?\rbrack\bot$.
Since $G^{\prime}$ $\models$ $\lbrack(\alpha_{G};\phi_{G}?;\vec{n};(\pi_{n_{1}}:=_{g}\bot);(\pi_{n_{1}}:=_{l}\top);\ldots;\vec{n};(\pi_{n_{k}}:=_{g}\bot);(\pi_{n_{k}}:=_{l}\top);\alpha_{1};\ldots;\alpha_{n})^{\star}\rbrack(\lbrack\alpha_{G};\phi_{G}?\rbrack\bot\rightarrow\varphi)$, then $G^{nf}$ $\models$ $\varphi$.
Thus, every normal form of $G^{\prime}$ with respect to $G\rightarrow(a_{1},\ldots,a_{n})$ satisfies $\varphi$.
\ \\
$\Rightarrow$: Suppose that every normal form of $G^{\prime}$ with respect to $G\rightarrow(a_{1},\ldots,a_{n})$ satisfies $\varphi$.
Let $G^{nf}$ be a finite rooted termgraph such that \\
$G^{\prime}$ $\longrightarrow_{(\alpha_{G};\phi_{G}?;\vec{n};(\pi_{n_{1}}:=_{g}\bot);(\pi_{n_{1}}:=_{l}\top);\ldots;\vec{n};(\pi_{n_{k}}:=_{g}\bot);(\pi_{n_{k}}:=_{l}\top);\alpha_{1};\ldots;\alpha_{n})^{\star}}$ $G^{nf}$ and $G^{nf}$ $\models$ $\lbrack\alpha_{G};\phi_{G}?\rbrack\bot$.
Then $G^{nf}$ is a normal form of $G^{\prime}$ with respect to $G\rightarrow(a_{1},\ldots,a_{n})$.
Hence, $G^{nf}$ satisfies $\varphi$.
Thus, $G^{\prime}$ $\models$ $\lbrack(\alpha_{G};\phi_{G}?;\vec{n};(\pi_{n_{1}}:=_{g}\bot);(\pi_{n_{1}}:=_{l}\top);\ldots;\vec{n};(\pi_{n_{k}}:=_{g}\bot);(\pi_{n_{k}}:=_{l}\top);\alpha_{1};\ldots;\alpha_{n})^{\star}\rbrack(\lbrack\alpha_{G};\phi_{G}?\rbrack\bot\rightarrow\varphi)$.
\end{proof}
\ \\
In other respects, the following proposition shows how an invariant $\varphi$
of a rewrite rule can be expressed in the proposed logic.
\begin{prop}\label{pro_5}
Let $\varphi$ be a modal formula.
The rewrite rule $G\rightarrow(a_{1},\ldots,a_{n})$ strongly preserves $\varphi$ iff $\models$ $\varphi\rightarrow\lbrack\alpha_{G};\phi_{G}?;\vec{n};(\pi_{n_{1}}:=_{g}\bot);(\pi_{n_{1}}:=_{l}\top);\ldots;\vec{n};(\pi_{n_{k}}:=_{g}\bot);(\pi_{n_{k}}:=_{l}\top);\alpha_{1};\ldots;\alpha_{n}\rbrack\varphi$.
\end{prop}
\begin{proof}
$\Leftarrow$: Suppose that $\models$ $\varphi\rightarrow\lbrack\alpha_{G};\phi_{G}?;\vec{n};(\pi_{n_{1}}:=_{g}\bot);(\pi_{n_{1}}:=_{l}\top);\ldots;\vec{n};(\pi_{n_{k}}:=_{g}\bot);(\pi_{n_{k}}:=_{l}\top);\alpha_{1};\ldots;\alpha_{n}\rbrack\varphi$.
Let $G^{\prime}$, $G^{\prime\prime}$ be finite rooted termgraphs such that $G^{\prime}$ $\models$ $\varphi$ and $G^{\prime}$ $\rightarrow_{G\rightarrow(a_{1},\ldots,a_{n})}$ $G^{\prime\prime}$.
Then $G^{\prime}$ $\models$ $\lbrack\alpha_{G};\phi_{G}?;\vec{n};(\pi_{n_{1}}:=_{g}\bot);(\pi_{n_{1}}:=_{l}\top);\ldots;\vec{n};(\pi_{n_{k}}:=_{g}\bot);(\pi_{n_{k}}:=_{l}\top);\alpha_{1};\ldots;\alpha_{n}\rbrack\varphi$ and \\
$G^{\prime}$ $\longrightarrow_{\alpha_{G};\phi_{G}?;\vec{n};(\pi_{n_{1}}:=_{g}\bot);(\pi_{n_{1}}:=_{l}\top);\ldots;\vec{n};(\pi_{n_{k}}:=_{g}\bot);(\pi_{n_{k}}:=_{l}\top);\alpha_{1};\ldots;\alpha_{n}}$ $G^{\prime\prime}$.
Hence, $G^{\prime\prime}$ $\models$ $\varphi$.
Thus, the rewrite rule $G\rightarrow(a_{1},\ldots,a_{n})$ strongly preserves $\varphi$.
\ \\
$\Rightarrow$: Suppose that the rewrite rule $G\rightarrow(a_{1},\ldots,a_{n})$ strongly preserves $\varphi$.
Let $G^{\prime}$, $G^{\prime\prime}$ be finite rooted termgraphs such that $G^{\prime}$ $\models$ $\varphi$ and \\
 $G^{\prime}$ $\longrightarrow_{\alpha_{G};\phi_{G}?;\vec{n};(\pi_{n_{1}}:=_{g}\bot);(\pi_{n_{1}}:=_{l}\top);\ldots;\vec{n};(\pi_{n_{k}}:=_{g}\bot);(\pi_{n_{k}}:=_{l}\top);\alpha_{1};\ldots;\alpha_{n}}$ $G^{\prime\prime}$.
Then $G^{\prime}$ $\rightarrow_{G\rightarrow(a_{1},\ldots,a_{n})}$ $G^{\prime\prime}$ and $G^{\prime\prime}$ $\models$ $\varphi$.
Thus, $\models$ $\varphi\rightarrow\lbrack\alpha_{G};\phi_{G}?;\vec{n};(\pi_{n_{1}}:=_{g}\bot);(\pi_{n_{1}}:=_{l}\top);\ldots;\vec{n};(\pi_{n_{k}}:=_{g}\bot);(\pi_{n_{k}}:=_{l}\top);\alpha_{1};\ldots;\alpha_{n}\rbrack\varphi$.
\end{proof}
\section{Conclusion}
\label{conclusion}

We have defined a modal logic which can be used either (i) to describe
data-structures which are possibly defined by means of pointers and
considered as termgraphs in this paper, (ii) to specify programs
defined as rewrite rules which process these data-structures or (iii)
to reason about data-structures themselves and about the behavior of
the considered programs. The features of the proposed logic are very
appealing.  They contribute to define a logic which captures
faithfully the behavior of termgraph rewrite systems. They also open
new perspectives for the verification of programs manipulating
pointers.

Our logic is undecidable in general. This is not surprising at all
with respect to its expressive power.  However, this logic is very
promising in developing new proof procedure regarding properties of
termgraph rewrite systems.  For instance, we have discussed a first
fragment of the logic, consisting of formulas without relabelling
actions, where validity is decidable.  Future work include mainly the
investigation of new decidable fragments of our logic and their
application to program verification.

\bibliographystyle{abbrv}
\bibliography{hylo}

\end{document}